\newlength{\extralineskip}
\documentclass[12pt]{article}

\usepackage{epsfig}
\begin{document}
\begin{titlepage}
\begin{flushright}
          \begin{minipage}[t]{12em}
          \large UAB--FT--465\\
                 April 1999
          \end{minipage}
\end{flushright}
\vspace{\fill}

\vspace{\fill}

\begin{center}
\baselineskip=2.5em

{\large \bf Casimir-Polder forces in the presence of the cosmic photon
heat bath}
\end{center}

\vspace{\fill}

\begin{center}
{\bf F. Ferrer and J.A. Grifols}\\
\vspace{0.4cm}
     {\em Grup de F\'\i sica Te\`orica and Institut de F\'\i sica
     d'Altes Energies\\
     Universitat Aut\`onoma de Barcelona\\
     08193 Bellaterra, Barcelona, Spain}
\end{center}
\vspace{\fill}

\begin{center}
\large Abstract
\end{center}
\begin{center}
\begin{minipage}[t]{36em}
We study the effect of a photon background at finite temperature $T$ on the
Van der Waals interactions among neutral bodies. It turns out that the
long-range Casimir-Polder force is unaffected for distances much less than
$T^{-1}$ and strongly enhanced for distances much above $T^{-1}$.
\end{minipage}
\end{center}

\vspace{\fill}

\end{titlepage}

\clearpage

\addtolength{\baselineskip}{\extralineskip}

The retarded dispersion potentials among neutral atoms or molecules have
been
studied since long time ago. Indeed, Casimir and Polder were the first
to obtain the
correct potential for the Van der Waals interaction of two neutral
systems with no
permanent electric dipole moment~\cite{cp}. Their potential differed at large
distances from the
non-relativistic Coulomb interaction derived by London~\cite{london}. These results
were obtained
using old fashioned perturbation theory and an important step forward
was the
introduction of modern quantum field
theory techniques. In this way a wide variety of phenomena associated to
dispersion forces was systematically studied~\cite{review}. Two-neutrino-exchange
forces, first
discussed by Feinberg and Sucher~\cite{neutri}, or spin independent forces arising
from double
(pseudo)scalar exchange~\cite{pseudo} provide examples of the activity in this field. 

 The asymptotic form for the Casimir-Polder potential, generalised in
ref.\cite{fs} to include magnetic effects, can be derived from the
phenomenological lagrangean density~\cite{itzyk}
\begin {equation}\label{lagrangean}
{\cal L}=-g_{1}\partial_{\alpha}\phi\partial^{\beta}\phi
F^{\alpha\gamma} F_{\beta\gamma}-g_{2}\phi^{2}F^2
\end {equation}
where $\phi$ is a scalar field, $F^{\alpha\beta}$ is the electromagnetic
field-strength, and 
$g_{1}={\alpha_{E}+\alpha_{B}\over 2m}$ and
$g_{2}=-{m \alpha_{B} \over 4}$. The relevant Feynman diagram is drawn
in Fig.1 and the resulting potential is
 
\begin {equation}\label{potential}
V_{CP}(r)=-
{\left[ 23(\alpha_{E}^{a}\alpha_{E}^{b}+\alpha_{B}^{a}\alpha_{B}^{b})-7(\alpha
_{E}^{a}\alpha_{B}^{b}+\alpha_{E}^{b}\alpha_{B}^{a}) \right] \over
(4\pi)^{3}r^{7}}.
\end {equation}      

\begin{figure}[bht]
\begin{center}
\epsfig{file=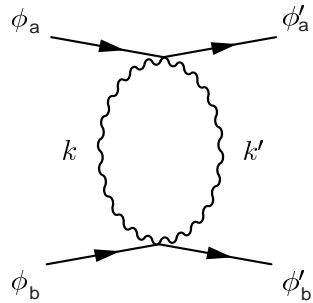,width=4cm,height=4cm}
\end{center}
\caption{\it Interaction between neutral systems arising from electric and
  magnetic susceptibilities.}
\end{figure}

In a photon populated medium, such as the cosmic microwave background
radiation (MWBR), one of the photons in the
double exchange can be supplied by the thermal bath. In the static limit,
i.e. momentum transfer $q \simeq(0, \vec Q)$, where matter is supposed to be at
rest in the microwave background radiation (MWBR) frame, the
potential is related to the Feynman amplitude in Fig.1 through
\begin{equation}\label{potential2}
V(r)={1\over(2\pi)^{2}r}\int {dQ}\,Q\sin Qr {i {\cal T}(Q)\over 4m_{a}m_{b}}.
\end{equation}
In this last formula, $Q \equiv | \vec Q |$ and ${\cal T} (Q) \equiv {\cal T}
(q=(0, \vec Q))$ with
\begin{equation} \label{amplitude}
{\cal T} (q)= \int {d^{4}k\,d^{4}k' \over (2\pi)^4} \delta^4(k+k'-q) d(k,k')
D_F (k^2) D_F(k'{}^2)
\end{equation}
where $g^{\mu\nu} D_F (k^2)$ stands for the photon propagator and the explicit
expression for the function $d(k,k')$ is~\cite{itzyk}:
\begin{eqnarray} \label{ditzykson}
d(k,k')=&8& g_{2}^{a}g_{2}^{b} K_{\mu\nu,\rho\sigma}(k) K^{\mu\nu,\rho\sigma}
(k') \nonumber \\
&+& \left[ 4g_{1}^{a}g_{2}^{b} (p_{\alpha}^{a}p_{\beta}'^{a} +
  p_{\beta}^{a}p_{\alpha}'^{a} ) K^{\alpha\nu,\rho\sigma}(k)
  K^{\beta}\,_{\nu,\rho\sigma}(k') + (a\leftrightarrow b) \right] \nonumber\\ 
&+& g_{1}^{a}g_{1}^{b} \left[ (p_{\alpha}^{a}p_{\beta}'^{a} +
  p_{\beta}^{a}p_{\alpha}'^{a}) K^{\alpha\nu,\lambda\kappa}(k)
  K^{\beta}\,_{\nu,}\,^{\delta}\,_{\kappa}(k') (p_{\lambda}^{b}p_{\delta}'^{b}
  + p_{\delta}^{b}p_{\lambda}'^{b}) + (k\leftrightarrow k') \right] 
\end{eqnarray}
with
$K_{\mu\nu,\rho\sigma}(k)=k_{\mu}k_{\rho}g_{\nu\sigma}-k_{\nu}k_{\rho}g_
{\mu\sigma}-k_{\mu}k_{\sigma}g_{\nu\rho}+k_{\nu}k_{\sigma}g_{\mu\rho}$.

The propagation of photons in a medium at temperature $T$ is described
by the Green function (in Feynman gauge)
\begin{equation}\label{propagator}
g^{\mu\nu} D_F(k,T)=-g^{\mu\nu} \left[ (k^2+i\epsilon)^{-1} - 2\pi
  i \delta (k^2) n(T,k^0) \right]
\end{equation}
where $n(T,k^0)$ is the Bose-Einstein distribution function for the
background photons\footnote{Following~\cite{HP}, to compute the
  $T$--dependent effects, we need only the real part of $i {\cal T} (Q)$
  correctly given by using~(\ref{propagator}), (see ref.~\cite{HP}), which is
  the $1-1$ component of the full $2$-dimensional matrix propagator used in the
  real time approach to finite temperature field theory~\cite{temp}.}. Of
course, use of the first piece 
in $D_F(k,T)$ above gives the zero temperature vacuum result in
equation~(\ref{potential}).
In a photon background, a contribution to the long range force can arise
because a photon in the thermal bath may be excited and de-excited back
to its
original state in the course of the double scattering process. This
effect is
described by the crossed terms contained in ${\cal T}(q)$ that involve the
thermal piece of one photon propagator along with the vacuum piece of
the
other photon propagator. This thermal component of the Feynman amplitude
can be written as
\begin{eqnarray} \label{thampl}
{\cal T}_{T}(q)=-i\int \!\! { d^{4}k \over(2\pi)^3}&& \left[ {1 \over
k^2} d(k,-(k-q)) n(T,k^0-q^0)\delta \left((k-q)^2 \right) \right. \nonumber \\
&& \left. + {1\over(k-q)^2}
d(k,-(k-q)) n(T,k^0) \delta(k^2) \right]
\end{eqnarray}     
where we used $\delta^4(k+k'-q)$ to integrate over $k'$. A shift of
variable $k-q\rightarrow k$ in the first
term of the sum in the previous expression leads to the more compact form, 
\begin{equation} \label{thampl2}
{\cal T}_{T}(q)=-i\int \!\!  { d^{4}k
\over(2\pi)^3} n(T,k^0) \delta(k^2) \left[ {d(k,-(k-q))\over(k-q)^2} +
{d(k+q,-k) \over(k+q)^2} \right].
\end{equation}
We integrate over the photon energy $\omega \equiv k^0$ with the help of
the Dirac delta and we are left with an integral
over three-momentum $\vec k$. The integral over the azimuthal angle in
$\vec k$-space is trivial and supplies a factor
$2\pi$. Hence, the thermal part of equation (3) can be written as,
\begin{equation} \label{potential3}
V_{T}(r) = { 1 \over 32m_{a}m_{b}\pi^{4}r } \int_{0}^{\infty} \!\! {\omega
\, d{\omega} \over e^{\omega/T}-1} \int _{-1}^{1} \!\! {dz} \,
\int_{0}^{\infty} \!\! {dQ} \sin Qr \left[ {d(k,-(k-q)) \over 2\omega
z-Q } - { d(k+q,-k) \over 2\omega z+Q} \right]_{k^0=\omega}.
\end {equation}
In the denominators of this formula we already took the static limit
$q\simeq(0,\vec Q)$. It is understood also that
the non relativistic limit for functions $d(k,-(k-q))$ and $d(k+q,-k)$
has been taken. That is, we set
$s=(p_{a}+p_{b})^2\simeq (m_{a}+m_{b})^2 $ and $t=(p_{a}-p'_{a})^2\simeq
-Q^2$, masses $m_{a},m_{b}$ being the
largest energy scales in the system, and use them to evaluate the scalar
products of momenta contained in (5). The process
is tedious but straightforward.  Let us now illustrate with some detail
the calculation of the component  of the
potential proportional to $g_{2}^{a}g_{2}^{b}$. The rest is calculated
along similar lines. The non-relativistic reduction
of the square bracket in (9) gives in this case:
\begin{equation} \label{bracket}
\left[ {d(k,-(k-q)) \over 2\omega z-Q } - { d(k+q, -k) \over 2\omega
z+Q} \right]_{k^0=\omega}^{(2,2)} = {128 Q^3 {\omega^2} z^2 \over
4 \omega^2 z^2 - Q^2}.
\end{equation}
Inspection of the potential (9) with the explicit form (10) in the
integrand shows that the integral over $Q$ is
ill-defined. It diverges for large $Q$. This is no surprise because the
effective lagrangean (1) is not renormalisable.
However, we are interested only on the long range (i.e. distances large
compared to the dimensions of the atoms)
behaviour of the potential and we may perform a finite number of short
distance subtractions without modifying the
asymptotic behaviour of the potential. For details of how this is done
in the vacuum case see~\cite{review,itzyk}. Here we
adopt the prescription $\int _{0}^{\infty}{dQ}\sin Qr{Q^3 \over (2\omega
z)^2-Q^2}\equiv -{d^2\over dr^2}\int
_{0}^{\infty}{dQ}\sin Qr{ Q  \over (2\omega z)^2-Q^2}$, where the
integral on the right is perfectly defined and gives
$-{\pi\over 2}\cos 2\omega zr$. This procedure will isolate the
asymptotic piece of the
potential. Hence, our potential reads, 
 
\begin{equation} \label{potential4}
V_{T}^{(2,2)}(r)={2g_{2}^{a}g_{2}^{b}\over m_{a}m_{b}\pi^{3}r}{d^2\over
dr^2}\int _{0}^{\infty}{\omega ^3 \,d{\omega}\over
e^{\omega/T}-1}\int _{-1}^{1}z^2\,{dz}\,\cos 2\omega zr.
\end{equation}
 The $z$-integral is easily done and
$V_{T}^{(2,2)}(r)$ is, 
\begin {equation}\label{potential5}
V_{T}^{(2,2)}(r)={g_{2}^{a}g_{2}^{b}\over m_{a}m_{b}\pi^{3}r}{d^2\over
dr^2} \left[ {1\over r^3} \left( -1+r{d\over dr}-{r^2\over
2}{d^2\over dr^2} \right) \int _{0}^{\infty}{ d{\omega}\over
e^{\omega/T}-1}\,\sin 2r\omega \right].
\end{equation}

The remaining thermal integral is found upon using the relation,
\begin {equation}\label{bose}
{1\over e^{\omega/T}-1}=\sum _{n=1}^{\infty}e^{-n \,\omega/T}
\end {equation}
 so that, once the $r$-derivatives are performed, the final
result boils down to  
\begin {equation}\label{potential6}
V_{T}^{(2,2)}(r)=\sum _{n=1}^{\infty}{3\alpha_{B}^{a}\alpha_{B}^{b}\over
8\pi^3r^7}(2rT)^6{[-n^4+10n^2(2rT)^2-5(2rT)^4]\over
[n^2+(2rT)^2]^5}
\end {equation}
where we reintroduced the magnetic polarisabilities of the particles $a$
and $b$.

Some more painful algebra incorporating all effects( electrical,
magnetic and mixed) leads to the total finite temperature
potential  
\begin{eqnarray}
\label{potential7}
V_{T}(r)=&-&{1\over 32\pi^3r^7} \left[ \left\{ 23( \alpha_{B}^{a}
   \alpha_{B}^{b} + \alpha_{E}^{a} \alpha_{E}^{b}) - 7 (\alpha_{E}^{a}
   \alpha_{B}^{b} + \alpha_{B}^{a} \alpha_{E}^{b}) \right\} {\cal S}_{0}
   (2rT)\nonumber \right.\\ 
&+& \left \{ 46( \alpha_{E}^{a} \alpha_{B}^{b} + \alpha_{B}^{a} \alpha_{E}^{b}
   ) - 14( \alpha_{B}^{a} \alpha_{B}^{b} + \alpha_{E}^{a} \alpha_{E}^{b})
   \right\} {\cal S}_{1}(2rT) \nonumber \\
&+& \left. \left\{ 11( \alpha_{B}^{a} \alpha_{B}^{b} + \alpha_{E}^{a}
   \alpha_{E}^{b}) + 5(\alpha_{E}^{a} \alpha_{B}^{b} + \alpha_{B}^{a}
   \alpha_{E}^{b}) \right\} {\cal S}_{2}(2rT) \right]
\end{eqnarray}
with ${\cal S}_{l}(x)\equiv x^{10-2l}\sum _{n=1}^{\infty}{n^{2l}\over
(n^2+x^2)^5}$.

We can sum over $n$ by realising that:
\begin {equation}\label{series}
\sum _{n=1}^{\infty}{1\over n^2+x^2}={1\over 2x^2}(\pi x\coth \pi x-1)
\end {equation}
and calculating $\sum _{n=1}^{\infty}{1\over (n^2+x^2)^5}$, $\sum
_{n=1}^{\infty}{n^2\over (n^2+x^2)^5}$ and
$\sum _{n=1}^{\infty}{n^4\over (n^2+x^2)^5}$ by differentiation of
equation~(\ref{series}) with respect to the parameter $x^2$
and by trivial algebraic manipulation. Once this is done we may obtain
analytic expressions for the functions ${\cal
S}_{l=0,1,2}(2rT)$ that enter the potential: 

\begin{eqnarray}
{\cal S}_0 (x) &=& \frac{1}{768} \left[ -384+105 x \pi+105 x \pi {\mathrm
    csch} (x \pi) \right. \nonumber \\
&&+ \left( 105 x^2 \pi^2+90 x^3 \pi^3+40 x^4 \pi^4+8 x^5 \pi^5 \right)
    {\mathrm csch}^2 (x \pi) \nonumber \\
&&+ \left( 90 x^3 \pi^3 + 24 x^5 \pi^5 \right) {\mathrm csch}^3 (x \pi)
    \nonumber \\ 
&&+ \left( 60 x^4 \pi^4 + 40 x^5 \pi^5 \right) {\mathrm csch}^4 (x \pi)
    \nonumber \\ 
&&+ \left. 24 x^5 \pi^5 {\mathrm csch}^5 (x \pi) \right] ,\\
&& \nonumber \\ 
{\cal S}_1 (x) &=& \frac{x \pi}{768} \left[15 + 15 {\mathrm csch} (x \pi)
 \right.    \nonumber \\  
&&+\left(15 x \pi+6 x^2 \pi^2-8 x^3 \pi^3-8 x^4 \pi^4 \right)
    {\mathrm csch}^2 (x \pi) \nonumber \\
&&+\left(6 x^2 \pi^2-24 x^4 \pi^4 \right)  {\mathrm csch}^3 (x \pi)\nonumber
    \\ 
&&+\left(-12 x^3 \pi^3-40 x^4 \pi^4 \right) {\mathrm csch}^4 (x \pi) \nonumber
    \\ 
&&-\left. 24 x^4 \pi^4 {\mathrm csch}^5 (x \pi) \right] 
\end{eqnarray}
and
\begin{eqnarray}
{\cal S}_2 (x) &=& \frac{x \pi}{768} \left[ 9+9 {\mathrm csch} (x \pi)
    \right. \nonumber \\  
&&+\left(9 x \pi-6 x^2 \pi^2-24 x^3 \pi^3+8 x^4 \pi^4 \right) {\mathrm csch}^2 (x \pi) \nonumber \\
&&+ \left(-6 x^2 \pi^2 + 24 x^4 \pi^4 \right) {\mathrm csch}^3 (x \pi)
    \nonumber \\
&&+\left(-36 x^3 \pi^3+40 x^4 \pi^4 \right)   {\mathrm csch}^4 (x \pi)
    \nonumber \\
&&+ \left. 24 x^4 \pi^4 {\mathrm csch}^5 (x \pi) \right] .
\end{eqnarray}

Now, our potential is only valid for distances much larger than the
dimensions of the objects involved. For the hydrogen
atom, for instance, one can estimate that distances should be on the
order or larger than $10^2\,\mbox{\AA}$~\cite{review,itzyk}. But the MWBR
temperature sets an additional distance scale:~$\sim (3K)^{-1}\sim
1\,mm$. So we may obtain approximate forms for the
potential in two limits, i.e.~$10^2\,\mbox{\AA} \le r \ll 1\,mm$ and~$ r\gg
1\,mm$.      

In the first case, for distances $r$ such that $rT\ll 1$ the  potential
is
\begin {equation}\label{potential8}
V_{T}(r)\simeq -{1\over 64
\pi^3r^7}\pi (rT) \left[ 32 ( \alpha_{B}^{a} \alpha_{B}^{b} + \alpha_{E}^{a}
\alpha_{E}^{b}) - 12( \alpha_{B}^{a} \alpha_{E}^{b} + \alpha_{E}^{a}
\alpha_{B}^{b}) \right].
\end {equation} 
On the other hand, when  $rT\gg 1$, we have
\begin {equation}\label{potential9}
V_{T}(r)\simeq -{1\over 64 \pi^3r^7} \pi (rT) 12 ( \alpha_{B}^{a}
\alpha_{B}^{b} + \alpha_{E}^{a} \alpha_{E}^{b} ).
\end {equation}
Obviously, in the $ r \ll 1\,mm$ regime, the effects on the original
Casimir-Polder potential due to the MWBR heat bath
are negligible and behaving as $r^{-6}$, but in the very large distance
domain, the dominant potential is the
temperature dependent one whose
$r$-behaviour is also $r^{-6}$. This is incidentally the behaviour of
the potential both in the old London model~\cite{london} and in
the Casimir-Polder approach at very short distances~\cite{cp}
(e.g. $1\,\mbox{\AA}<r\ll 10^2\,\mbox{\AA}$, in the hydrogen case) which we did not
bother to consider when displaying the asymptotic potential in equation
(2). For the intermediate region ($r\sim {\cal
O}(1\,mm)$), where both potentials are present with comparable
strengths, we may plot the exact result instead of 
writing down the explicit form of equation (15). This is conveniently
shown in Fig.2 where we display $V_{T}(r)/V_{CP}(r)$ as a function of
distance. We see clearly how, in the region of
interest, the curve interpolates between a steep linear behaviour on the
small $r$ side and again a linearly growing
function on the large side of $r$.

\begin{figure}[bht]
\begin{center}
\epsfig{file=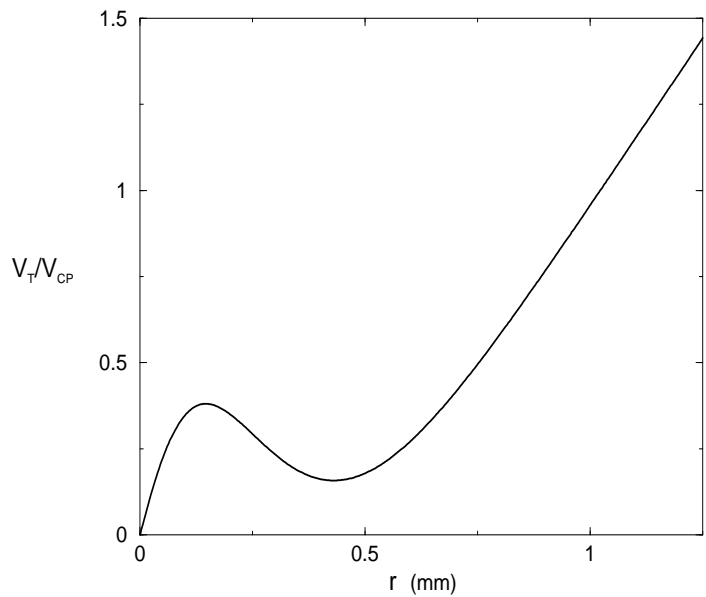,width=9cm,height=8cm}
\end{center}
\caption{\it Ratio between the potential $V_T$, at $T \sim 2.72
  K$ corresponding to the cosmic MWBR, and the zero temperature Casimir-Polder
  potential $V_{CP}$ when $\alpha_B^{a,b} \gg \alpha_E^{a,b}$ as e.g. for
  two H atoms.}
\end{figure}

We would like to close this paper with a few comments. Firstly, from the
theoretical standpoint, the effect of the MWBR
is definitely there: the Casimir-Polder force among neutral bodies gets
an additional contribution because these bodies sit
in a background of cosmic photons. We feel that the effect should be
calculated although, admittedly, the strength of
this force is far beyond present experimental capabilities. Indeed,
generic potentials with a power-law fall-off $r^{-n}$
have been experimentally scrutinised in the laboratory at various
distance scales. Even for the shortest ranges explored,
the limits on the strength of potentials with $n\ge 5$ are very
poor~\cite{mostepanenko}. Finally, we would like to point out that when
searching for extra forces in the
laboratory (like the ones hypothesised in different
completions of the Standard Model~\cite{price}) one has to be aware of the  real effects that are present in
the system. The Casimir-Polder force exists
and is good to know that it is not modified deep in the sub-millimeter
domain. 

Work partially supported by the CICYT Research Project
AEN98-1093. F.F. acknowledges the CIRIT for financial support.

\end{document}